# An Intelligent Approach for Dense Urban Area in existing 2G / 2.5G

Sumit Katiyar, Prof. R. K. Jain, Prof. N. K. Agrawal

**Abstract**— In the prevailing scenario audio, video, data services (i.e. internet), multimedia and broadcasting etc. are being integrated. Decreasing cell size increases capacity but at the same time increases fluctuation and interference too. The intelligence approach is the only answer in developing countries where frequency and power are scarce resources. In this paper, we have tried to integrate all proven technologies in networking such as in-building network, micro zone, intelligent micro cell, deployment along city streets, tunnels, subway coverage etc. along with adaptive frequency allocation in hierarchical approach with the help of adaptive / intelligent antenna system. A-SDMA approach will further enhance spectral efficiency as well as QoS (Quality of Service). It can be proved beyond doubt that this integrated approach will convert 2G / 2.5G systems capable of handling the prevailing demand at reduced cost. In addition to it, integrated approach will save power and reduce RF pollution. In this paper we have explained the ill effect of cellular growth in terms of health hazard and increased power consumption. We have also suggested ways and means to overcome these problems (spectral density / capacity, QoS, power consumption and RF pollution etc.).

——————————  ◆  ——————————

## 1 INTRODUCTION

AS the mobile telecommunication systems are tremendously growing allover the world then the numbers of handheld and base stations are also rapidly growing and it became very popular to see these base stations distributed everywhere in the neighborhood and on roof tops which has caused a considerable amount of panic to the public in INDIA concerning wither the radiated electromagnetic field from these base stations may cause any health effect or hazard. The radiated electromagnetic energy from mobile base stations and the exposure level due to these stations were measured and compared to some international standards and guidelines in order to answer some of the public fear and concern by A. Mousa [1]. Due to practical constraints the shape of the cells are not perfect circles or hexagonal but it depends on the environments such as buildings, mountains, trees, rivers, traffic density and other miscellaneous factors, it also depends on weather conditions and even system load.

The provision of capacity for the increasing traffic demand in mobile radio networks comes along with the reduction of the cell size and, hence, stronger traffic fluctuations between the cells. Moreover, improved indoor coverage is required. Hierarchical cellular structures can serve indoor users and hot spots by pico- and micro cell layers, respectively, while providing coverage in the area by the macro cell layer. Moreover, hierarchical cellular structures can compensate traffic fluctuations e.g. by shifting overflow traffic from lower to higher layers. In order to avoid inter ference between the layers, their frequency allocations have to be coordinated. This can be achieved by incorporating smart antenna / intelligent antenna in hierarchical structure with adaptive-SDMA approach.

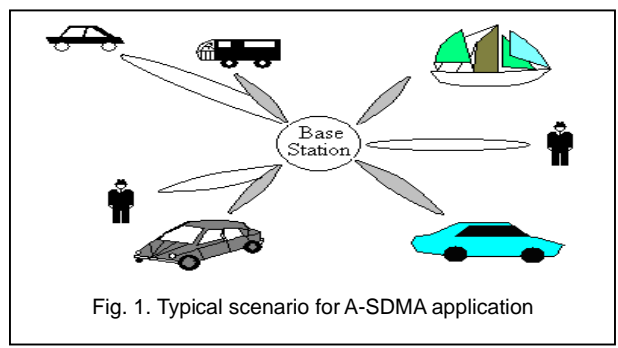

Fig. 1. Typical scenario for A-SDMA application

For mobile applications, the directivity of the antennas has to be adapted such that the beams pointing to the users can follow their movements. This is called adaptive SDMA or A-SDMA. An example of a typical A-SDMA scenario is depicted in Fig. 1, where seven mobile stations are served on only two carrier frequencies marked by bright and dark beams, respectively [27].

Due to small size cells, interference (co-channel & adjacent channel) increases and it becomes difficult to maintain SIR. This has resulted in unplanned growth of base stations in INDIA. Vendors are not using the simple required technologies for handling this increased traffic but they are using shortcuts and this is the reason for more RF pollution and increased power consumption. Vendors are using umbrella structure (macro, micro cells). Micro cells should be operated at low power but vendors are using full power micro cells for penetrating signals for indoor subscribers instead of using simple in-building networks in INDIA. This paper focuses on the ill effect of RF exposure to man kind and the simple available technologies to overcome this problem.

————————————

- Sumit Katiyar is currently pursuing PhD degree program in electronics engineering from Singhania University, India, E-mail: sumitkatiyar@gmail.com
- Prof. R. K. Jain is currently pursuing PhD degree program in electronics engineering from Singhania University, India,
  E-mail: rkjain_iti@rediffmail.com
- Prof. N. K. Agrawal is a senior member of IEEE and life member of ISTE and ISCEE. E-mail: agrawalnawal@gmail.com

## 2 ELECTROMAGNETIC RADIATION AND SAFETY ISSUES OF CELLULAR COMMUNICATION

Electromagnetic radiation may be classified as ionizing and non-ionizing radiation. Ionizing radiation has enough energy to remove bound electrons from the orbit of an atom such that it becomes an ionized atom which may cause health hazard. On the other hand, the non-ionizing radiation does not have the sufficient energy to ionize (change) the atoms. For example, the human eye can easily perceive the light whereas the EMF with very high frequency, like X-ray, may ionize material and break down molecules. However, this radioactive radiation should not be confused with radio wave frequency (RF).

The radio waves used in mobile telephones and cellular communications are also electromagnetic waves like visible light and X-ray and they also propagate in the same speed of light. The RF used for mobile communication can be in the range 450-2200 MHz which is considered as part of the microwave (MW) range as shown in Figure 2 [1].

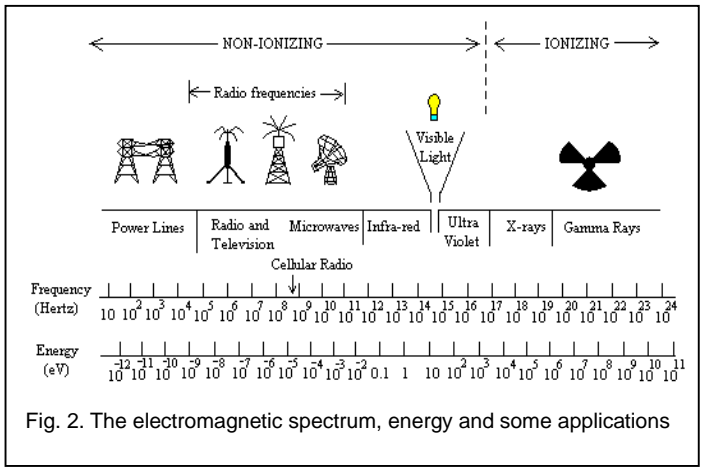

Fig. 2. The electromagnetic spectrum, energy and some applications

## 3 CELLULAR NETWORKS

The cellular communication industry is arising in INDIA with fastest pace. It consist of more than 4 lacs base stations (which are increasing with fast rate) covering adjoining zones, called size and the used multi-million mobile phones. We adopted the following cellular networks in INDIA. Table 1 & 2 illustrates the basic parameters of cellular networks in INDIA.

Mobile phones are connected to the base stations via radio frequency electromagnetic waves and vice-versa. These mobile phones should be in the coverage area of a certain cell; the base station of this cell must assign this mobile phone the necessary RF channel to maintain the connection. To achieve this link, the mobile should use certain transmission power to con-nect to the base station (uplink) and the base station should use the required power to connect to the mobile phone (down link).

TABLE 1
GSM FREQUENCY SPECTRUM

| GSM range (MHz) | Uplink (MHz) | Downlink (MHz) |
|---|---|---|
| 900 | 890-915 | 935-960 |
| 1800 | 1710-1785 | 1805-1880 |
| 1900 | 1850-1910 | 1930-1990 |

TABLE 2
MOBILE RADIO SYSTEMS IN INDIA

| Cellular System | Year | Transmission Type | Multiple Access Type | Channel Bandwidth | Generation |
|---|---|---|---|---|---|
| US Narrow band Spectrum !S-95, CDMA one | 1993 | Digital | CDMA | 800 / 1900 MHz | $2^{nd}$ |
| CDMA-200 | 2001 | Digital | CDMA | !900 MHz | $3^{rd}$ |
| Global System of Mobile GSM | 1900 | Digital | TDMA | 900/ 1800 MHz | $2^{nd}$ |
| Universal Mobile Telecom System UMTS WCDMA | 2001 | Digital | CDMA | 2000 MHz | $2^{nd}$ |

It is important to conserve the battery life and to minimize the interference between cells and hence the mobile transmission power is limited to the minimum level required to maintain the connection and so it is better to be close to the base station so as to reduces the transmission power in both directions.

Reduction in power consumption will automatically result in reduction in RF pollution level, interference level. This will also result in enhancement of\ spectral density & QoS, which is the need of hour in INDIA. In this paper we have suggested the network which will be based on above concept.

### 3.1 Safety Guidelines to Level of Exposure

Since it was first introduced around 1980, almost all classes of society are using mobile phones. The estimated number of mobile phone users is around multi billion all over the world [2]. The Cellular Telecommunication Industry Association (CTIA) has reported that this industry is growing at a rate of 40% per year [3]. This so fast rapid deployment of mobile telecommunications networks in the world has alert public concern over possible health effects from exposure to radiofrequency electromagnetic energy (RF EME) emitted by both mobile phones and base station antennas.

1. Health and Safety- The exposure to very high levels of RF energy can be harmful and may cause some biological effects on human [4], [5], these effects may result from heating of tissues and it is called "thermal effects" this is mainly due to the ability of RF energy to heat biological tissue, this is the same as the principle of the microwave oven. On the other hand, at low levels of exposure to RF radiation (lower than the threshold which may cause heat), the evidence of harmful biological effects is unpro-ven; however it may cause some non thermal effects [6]. At the extremely high frequencies like X-rays, electromagnetic particles have sufficient energy to break bonds (ionization). This is why genetic material of cells may be damaged leading to cancer or birth defects. However, at lower frequencies, such as the RF ones used by cellular communications, the energy of the particles is too much low to break chemical bonds. Thus, RF energy is non-ionizing [7]. This RF (radio frequency) energy causes neurological, cardiac, respiratory, ophthalmological, dermatological and other conditions ranging in severity from headaches, fatigue and ADD to pneumonia, psychosis and strokes.

2. Standards and Limitations- The RF exposure guidelines are developed for both the general public (unconditional exposure) and those who are working with this field's occupational exposure by many organizations and endorsed by WHO too. For the case where the exposure is local (RF is transmitted from the mobile phone and hence closer to the user), the highest power absorption per unit mass in a small part of the body must be used and compared with the recommendations and standards. Specific Absorption Rate (SAR) is the quantity used to measure this amount of RF energy and it is expressed in units of watts per kilogram (W/kg) or (mW/kg). The maximum permissible exposure MPE recommended for power density is based on the threshold SAR value. This data has been illustrated by many organizations and safety aspects have been evolved accordingly [1], [6], [8], [9].

### 3.2 Basic Technologies
Basic technologies have been chosen on the following concept:
1. Use of appropriate power levels taking care of environmental condition and system load.
2. QoS
3. SIR- Required SIR is maintained through reduction in interference (using minimum signal power) for controlling RF pollution.
4. Spectral Density

## 4 HIERARCHICAL SYSTEM
The demand for cellular radio services growing rapidly, and in heavy populated areas the need arises to shrink the cell sizes and scale the testing pattern. The extension of the service into the PCN domain, railway stations, malls, pedestrian areas, markets and other hotspots further enhances this trend. The vision of future cellular systems incorporates macro, micro, pico and femto cells in hierarchical structure.

### 4.1 Macro Cell
A conventional base station with 20W power and range is about 1 km to 20 km.

### 4.2 Micro Cell
A conventional base station with 5W power and range is about 500 m to 2 km. Microcells and picocells take care of slow traffic (pedestrian and in-building subscribers) (Fig 3).

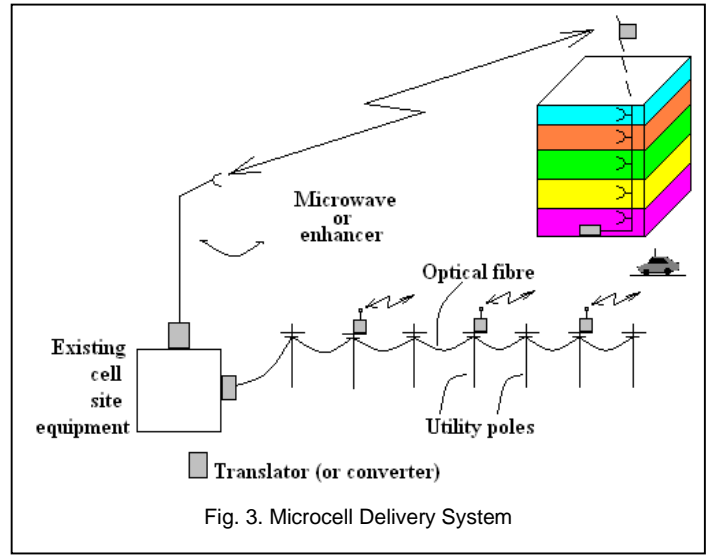

Fig. 3. Microcell Delivery System

The picocells are small versions of base stations, ranging in size from a laptop computer to a suitcase. Besides plugging coverage holes, picocells are frequently used to add voice and data capacity, something that repeater and distributed antenna can not do.

Picocells are designed to serve very small area such as part of a building, a street corner, malls, railway station etc.

### 4.4 Femto Cell
A femtocell is a smaller base station, typically designed for use in home or small business. It connects to the service provider's network via broadband (such as DSL or cable); current designs typically support 2 to 4 active mobile phones in a residential setting, and 8 to 16 active mobile phones in enterprise settings. A femtocell allows service providers to extend service coverage indoors, especially where access would otherwise be limited or unavailable.

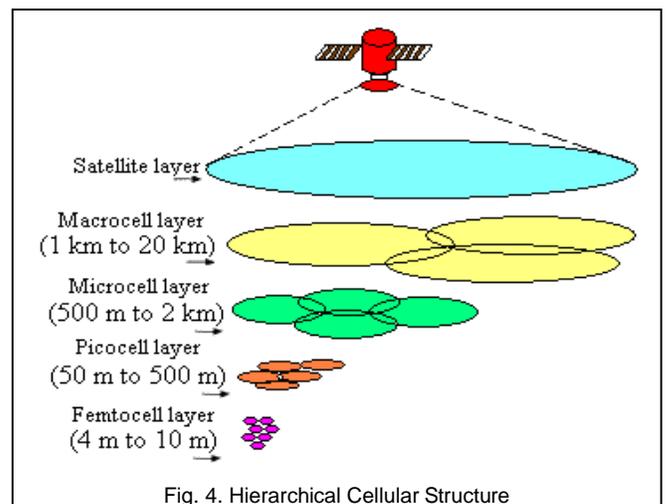

Fig. 4. Hierarchical Cellular Structure

This hierarchical cellular structure gives better coverage, improved capacity, improved macro / micro cell reliability, reduced cost, reduced power consumption and RF pollution.

## 5 HANDOFF IN CELLULAR SYSTEM

Handoff is an essential element of cellular communications. Efficient handoff algorithms are a cost-effective way of enhancing the capacity and QoS of cellular systems. Macrocell radii are in several kilometers. Due to the low cell crossing rate, centralized handoff is possible despite the large numbers of MSs are to be managed by MSC. The use of microcells is considered the single most effective means of increasing the capacity of cellular systems. Microcells are more sensitive to the traffic and interference than macrocells due to short term variations (e.g., traffic and interference variations), medium / long term variations (e.g., new buildings), and incremental growth of the radio network (e.g., new BSs) [11]. The number of handoffs per cell is increased by an order of magnitude, and the time available to make a handoff is decreased [12]. Using an umbrella cell is one way to reduce the handoff rate. Due to the increase in the microcell boundary crossings and expected high traffic loads, a higher degree of decentralization of the handoff process becomes necessary [13].

Lot of handoff schemes has been developed. The schemes packages show that these schemes can significantly decrease both the number of dropped handoff calls and the number of blocked calls without degrading the quality of communication service and the soft handoff process [11], [12], [13], [14]

## 6 OFDMA

In an OFDMA network, each user transmits OFDM symbols by using a group of subcarriers exclusive from other users. As a result, multiple users can transmit simultaneously without interfering with each other. OFDMA also inherits the advantages of the OFDM technique, such as high spectral efficiency and resistance to multipath fading. However, when the subcarrier allocation is fixed, the performance of a user may suffer severely if its allotted subcarriers experience deep fade. In a multi-user OFDMA system, owing to the independence of fading among different users, the subcarriers and power allocation, the system performance can be improved significantly. The authors consider multi-access control for the uplink in OFDMA wireless networks, where subcarriers are grouped into clusters [21]. In addition, OFDM technology possesses a number of unique features that makes it an attractive choice for high speed broadband wireless communications:

1. OFDM is robust against multipath fading and inter-symbol interference because the symbol duration increases for the lower rate parallel subcarriers (For a given delay spread, the implementation complexity of an OFDM receiver is considerably simpler than that of a single carrier with an equalizer).
2. OFDM allows for an efficient use of the available radio frequency (RF) spectrum through the use of adaptive modulation and power allocation across the subcarriers that are matched to slowly varying channel conditions using programmable digital signal processors, thereby enabling bandwidth on demand technology and higher spectral efficiency.

OFDM makes single frequency networks possible, which is particularly attractive for broadcasting applications [22].

## 7 SMART / ADAPTIVE ANTENNA

A smart antenna is an array of antenna elements connected to a digital signal processor. Such a configuration (Fig 5) dramatically enhances the capacity of a wireless link through a combination of diversity gain, array gain, and interference suppression (Fig 6 (b)).

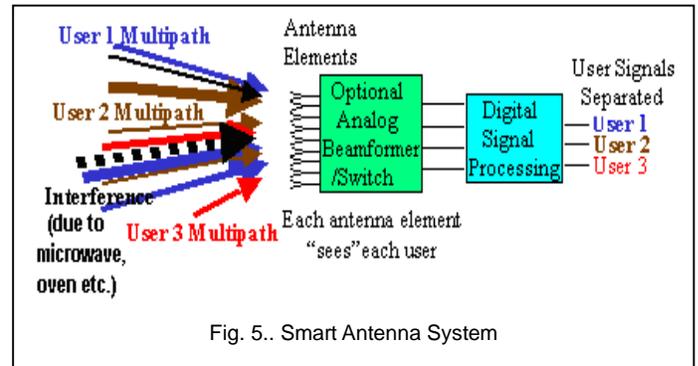

Fig. 5.. Smart Antenna System

Each antenna element "sees" each propagation path differently, enabling the collection of elements to distinguish individual paths to within a certain resolution. As a consequence, smart antenna transmitters can encode independent streams of data onto different paths or linear combinations of paths, thereby increasing the data rate, or they can encode data redundantly onto paths that fade independently to protect the receiver from catastrophic signal fades, thereby providing diversity gain. A smart antenna receiver can decode the data from a smart antenna transmitter this is the highest-performing configuration or it can simply provide array gain or diversity gain to the desired signals transmitted from conventional transmitters and suppress the interference [15], [16].

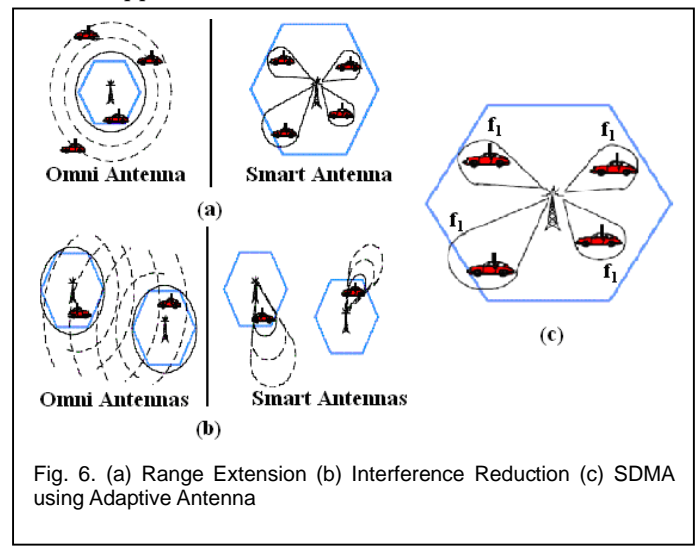

Fig. 6. (a) Range Extension (b) Interference Reduction (c) SDMA using Adaptive Antenna

Furthermore, smart antennas provide maximum flexibility by enabling wireless network operators to change antenna patterns to adjust to the changing traffic or RF conditions in the network [17].

Smart antennas at base stations can be used to enhance mobile communication systems in several ways:

1. Increased BS range (Fig 6 (a))
2. Less interference within the cell

3. Less interference in neighboring cells
4. Increased capacity by means of SFIR or SDMA

'Smart' antenna transmitters emit less interference by only sending RF power in the desired directions. Furthermore, 'smart' antenna receivers can reject interference by looking only in the direction of the desired source. Consequently 'smart' antennas are capable of decreasing CCI. A significantly reduced CCI can be taken as advantage of Spatial Division Multiple Access (SDMA) [18]. The same frequency band can be re-used in more cells and the same frequency can be used in same cell for different subscribers (Fig 6 (c)), i.e. the so called frequency re-use distance can be decreased. This technique is called Channel Re-use via Spatial Separation. In essence, the scheme can adapt the frequency allocations to where the most users are located [19]. With the inclusion of nanotechnology devices, the capability of adaptive antenna will increase manifold.

## 8 PROPOSED NETWORK

Proposed network (Fig 7) is based on umbrella structure where in macro cell will be used for fast traffic and micro cell will be used for slow traffic. Adaptive / Intelligent antenna will be used for estimation of user's velocity. As user estimation of user velocity is important for efficient system management in mobile cellular systems [23]. The RF resources will be dynamically allocated between macro and micro cell accordingly (Adaptive Frequency Allocation scheme will be followed) [24]. It is also assumed that the MS is equipped with a rake receiver capable of performing "maximal ratio combining" of the signals it receives from the transmitting BSs [25].

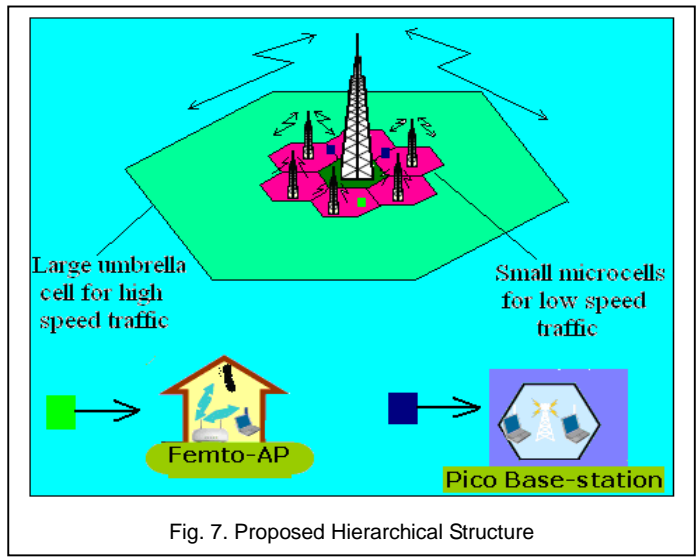

Fig. 7. Proposed Hierarchical Structure

Pico cell will be marked for hotspots. An adaptive frequency allocation will be applied as strategy to reuse the macro and micro cell resources without frequency planning in indoor picocells [26].

Femto cell will be marked as home base stations- which are data access points installed by home users to get better indoor voice and data coverage. Femtocells enable a reduced transmit power, while maintaining good indoor coverage. Penetration losses insulate the femtocell from surrounding femtocell transmissions. As femtocells serve only around 1-4 users, they can devote a larger portion of their resources (transmit power & bandwidth) to each subscriber.

The above proposed structure will undoubtedly enhance the spectral density with the help of diversity and adaptive approach through rake receiver and adaptive antenna respectively. The induction of pico and femto cell will reuse the RF resources of overlaid macro / micro structures which will enhance spectral density manifold. The simple technologies suggested by William C Y Lee for deployment along city streets, deployment along binding roads, deployment under the ground (subway coverage) and in-building designs etc. will also be considered in proposed hierarchical structure [20].

## 9 DISCUSSION

Adaptive / Intelligent antennas combined with SDMA approach not only suppress but also achieve adaptive antenna pattern which can adapt the frequency allocation to where the most users are located. With the inclusion of nanotechnology devices the capabilities of adaptive / intelligent antenna will increase manifold. This has been proved beyond doubt that spectral density will be enhanced substantially. In addition to it overlaid structures will reduce the handoff using simple handoff algorithms. Moreover RF resources of macro / micro will be reused in pico cell without frequency planning. This will further enhance the capacity of the system beyond any doubt. The use of femto cell will further increase the capacity of the system. In essence proposed network will be a highly efficient cost-effective way of enhancing the capacity and QoS of cellular systems. The above proposed system will be highly effective energy saving cellular network which will reduce power consumption and RF pollution significantly.

## 10 CONCLUSION

In this paper the ill effect of RF pollution due to deployment of unplanned cellular network in urban areas has been addressed. A simple network based on proved technologies [20] has been recommended. More emphasis is given on reduction of interference for maintaining desired SIR and QoS in high density and complex environment. In umbrella structure adaptive frequency allocation approach has been followed for maximal utilization of RF resources between micro and macro cells with the help of adaptive / intelligent antenna clubbed with SDMA. These macro / micro cell resources have been reused without frequency planning in pico cells and very low power femto cells have been added for individuals to take care of internet users and small corporate. It is clear beyond doubt that proposed network will enhance spectral density along with reduction of interference as well as RF pollution and reduce power consumption. As a result, this network will able to control RF pollution with in limits and will not surpass the level which will be harmful to mankind / animals.